\documentclass[conference]{IEEEtran}
\IEEEoverridecommandlockouts
\usepackage{cite}
\usepackage{amsmath,amssymb,amsfonts}
\usepackage{algorithmic}
\usepackage{float}
\usepackage{graphicx}
\usepackage{textcomp}
\usepackage{xcolor}
\usepackage{multirow}
\usepackage{booktabs}
\usepackage{array}
\usepackage{tabularx}
\usepackage{longtable}
\usepackage{makecell}
\usepackage{etoolbox}
\usepackage{mathtools}
\usepackage{xurl}
\usepackage[hidelinks,breaklinks=true]{hyperref}

\makeatletter
\def\@IEEEaftertitlespace{0pt}
\makeatother

\BeforeBeginEnvironment{tabbing}{\vspace{-4pt}\par\noindent}
\AfterEndEnvironment{tabbing}{\vspace{-6pt}}

\begin{document}
\title{\vspace{-8mm}Data Center Spatio-Temporal Load Flexibility in Security-Constrained Unit Commitment for Enhanced Grid Efficiency and Reliability\vspace{-4mm}}
\author{\IEEEauthorblockN{Haoxiang Wan}
\IEEEauthorblockA{\textit{Department of Electrical and Computer Engineering}\\
\textit{University of Houston}, Houston, TX, USA\\
hwan6@cougarnet.uh.edu}
\and
\IEEEauthorblockN{Xingpeng Li}
\IEEEauthorblockA{\textit{Department of Electrical and Computer Engineering}\\
\textit{University of Houston}, Houston, TX, USA\\
xli83@central.uh.edu}
\vspace{-16mm}
}
\maketitle
\begin{abstract}
Data center electricity consumption reached 4.4\% of U.S. total in 2023 and is projected to grow to 6.7--12\% by 2028, imposing increasing stress on transmission networks while representing a largely untapped source of controllable demand-side flexibility. This paper proposes a modular security-constrained unit commitment (SCUC) framework that coordinates flexible data center workloads with system-level scheduling to reduce renewable curtailment, alleviate congestion, and lower operating costs. Three mixed-integer linear programming (MILP) models are formulated: the Data Center Spatial model (DC-S), enabling instantaneous workload redistribution across geographically distributed sites; the Data Center Temporal model (DC-T), permitting each site to shift its deferrable load across time while preserving the daily energy balance; and the Data Center Spatio-Temporal model (DC-ST), jointly activating both mechanisms and spanning the largest feasible operating region. Case studies on a modified IEEE 24-bus reliability test system show that DC-ST eliminates all base-case and post-contingency transmission violations at a flexibility ratio of 40\%, and reduces renewable curtailment by up to 84.4\% at 30\% relative to the inflexible baseline. Sensitivity analysis further reveals that moderate flexibility levels of 20\%--30\% already capture most of the achievable benefits, supporting practical deployment with limited operational burden on data center operators.
\end{abstract}

\vspace{+4pt}
\vspace{-4pt}
\section*{Nomenclature}
\vspace{-2pt}
{\fontsize{8}{9.5}\selectfont
\noindent\textit{Sets and Indices:}
\begin{tabbing}
\hspace{2.2cm}\= \kill
$\mathcal{N}$ \> Set of all buses. \\
$\mathcal{G}$ \> Set of all generators. \\
$\mathcal{G}_n$ \> Set of generators at bus $n$. \\
$\mathcal{L}$ \> Set of all branches. \\
$\mathcal{T}$ \> Set of all time periods, $\{1,\ldots,24\}$. \\
$\mathcal{K}$ \> Set of \textit{N}-1 contingency scenarios. \\
$\mathcal{D}$ \> Set of data center buses. \\
$\mathcal{S}$ \> Set of renewable generation buses. \\
$\delta^{+}(n)$/$\delta^{-}(n)$ \> Receiving-/sending-end branch sets at bus $n$. \\
$g$ \> Generator index. \\
$n$ \> Bus index. \\
$l$ \> Branch index. \\
$t$ \> Time period index. \\
$c$ \> Contingency scenario index. \\
$f_l$ / $t_l$ \> Sending-/receiving-end bus of branch $l$.
\end{tabbing}
\noindent\textit{Generator Parameters:}
\begin{tabbing}
\hspace{2.2cm}\= \kill
$C^{\mathrm{E}}_g$ \> Linear variable energy cost (\$/MWh). \\
$C^{\mathrm{NL}}_g$ \> No-load cost (\$). \\
$C^{\mathrm{SU}}_g$ \> Start-up cost (\$). \\
$P^{\min}_g$/$P^{\max}_g$ \> Minimum/maximum output limits (p.u.). \\
$R^{\mathrm{SP}}_g$ \> Min spinning reserve ramp rate (p.u.). \\
$R^{\mathrm{H}}_g$ \> Hourly ramp rate (p.u./h). \\
$R^{\mathrm{SR}}_g$ \> Start-up ramp capability (p.u.). \\
$R^{\mathrm{SD}}_g$ \> Shut-down ramping limit of generator $g$. \\
$T^{\mathrm{up}}_g$/$T^{\mathrm{dn}}_g$ \> Minimum up/down times (h).
\end{tabbing}
\noindent\textit{Data Center Parameters:}
\begin{tabbing}
\hspace{2.2cm}\= \kill
$D^{\mathrm{orig}}_t$ \> Pre-optimization data center load profile (p.u.). \\
$\overline{D}_n$ \> Physical capacity of site $n$ (p.u.). \\
$\beta$ \> Flexibility ratio, $\beta \in [0,1]$. \\
$D^{\mathrm{fix}}_{n,t}$ \> Non-deferrable load component (p.u.) \\
$\hat{D}^{\mathrm{flex}}_{n,t}$ \> Nominal flexible quota (p.u.).
\end{tabbing}
\noindent\textit{Decision Variables:}
\begin{tabbing}
\hspace{2.5cm}\= \parbox[t]{5.5cm}{x}\kill
$P_{g,t}$ \> Generator output (p.u.). \\
$u_{g,t}$ \> Commitment status (binary). \\
$v_{g,t}$ \> Start-up indicator (binary). \\
$R^{v}_{g,t}$ \> Spinning reserve (p.u.). \\
$\theta_{n,t}$ \> Voltage angle at bus $n$ (rad). \\
$P^{f}_{l,t}$ \> Branch flow on line $l$ (p.u.). \\
$D_{n,t}$ \> \parbox[t]{5.5cm}{Total data center (DC) load at site $n$, co-optimized with unit commitment (p.u.).} \\[6pt]
$D^{\mathrm{flex}}_{n,t}$ \> \parbox[t]{5.5cm}{Optimized flexible load component, distinct from nominal quota $\hat{D}^{\mathrm{flex}}_{n,t}$ (p.u.).} \\[6pt]
$R^{\mathrm{use}}_{n,t}$ \> Utilized renewable output (p.u.). \\
$R^{\mathrm{curt}}_{n,t}$ \> Curtailed renewable output (p.u.). \\
$\alpha_{l,t}$ \> Base-case thermal violation slack (p.u.). \\
$\sigma_{n,t}$ \> Base-case load shedding (p.u.). \\
$P^{c}_{g,t}$ \> Post-contingency generator output (p.u.). \\
$\theta^{c}_{n,t}$ \> Post-contingency voltage angle (rad). \\
$P^{f,c}_{l,t}$ \> Post-contingency branch flow (p.u.). \\
$\alpha^{c}_{l,t}$ \> Post-contingency thermal violation slack (p.u.). \\
$\sigma^{c}_{n,t}$ \> Post-contingency load shedding (p.u.).
\end{tabbing}}
\section{Introduction}
\vspace{-4pt}
The rapid expansion of cloud computing, artificial intelligence, and digital services has driven sustained growth in large-scale data center (DC) electricity demand \cite{b1,b2}. According to the 2024 U.S. Department of Energy report, U.S. DC electricity consumption reached 4.4\% of the national total in 2023 and is projected to grow to 6.7--12\% by 2028 \cite{b23}, reshaping regional load profiles and introducing new operational challenges for power system operators, including heightened transmission congestion, increased reliance on high-cost peaking resources, and reduced scheduling flexibility \cite{b4,b5,b6}. Concurrently, the large-scale integration of renewable energy like solar photovoltaic (PV) generation has amplified system uncertainty; its location-dependent, time-concentrated output can exacerbate network congestion and, when delivery to load centers is constrained, induce renewable curtailment \cite{b3,b5,b6}.

Unlike conventional industrial loads, DCs serve a heterogeneous mix of latency-critical and delay-tolerant workloads \cite{b7,b20,b21}. Computationally intensive but scheduling-flexible tasks---batch processing, model training, large-scale analytics---can be deferred in time or, for operators with geographically distributed facilities, migrated across sites. This characteristic positions DCs as a structurally unique demand-side resource: one that can be coordinated to absorb renewable generation during high-output intervals, relieve transmission bottlenecks, and reduce reliance on costly corrective actions \cite{b7,b8,b20}.

Prior work on DC-based demand response and workload management has focused predominantly on the operator's perspective, targeting electricity cost minimization, carbon-aware scheduling, and geographic load balancing \cite{b22,b11,b10}. While these studies establish the technical feasibility and economic incentives for flexible computing, the integration of DC flexibility into power system scheduling---particularly within security-constrained operational frameworks subject to unit commitment logic and \textit{N}-1 contingency requirements---remains comparatively underexplored. Furthermore, the spatial and temporal dimensions of DC flexibility are typically studied in isolation, obscuring their complementary operational roles and precluding a direct quantification of the marginal value of each dimension.

Motivated by these gaps, this paper develops a security-constrained unit commitment (SCUC) framework that co-optimizes generator scheduling with flexible DC demand as a system-level resource, extending prior SCUC formulations focused on computational acceleration, corrective redispatch, and conventional demand response \cite{b12,b13,b18}. Workload flexibility is captured through modular constraints representing (i) spatial workload redistribution across DC sites, (ii) temporal load shifting at the same site, and (iii) their joint activation, each maintaining rigorous energy accounting and full compatibility with power flow and \textit{N}-1 contingency modeling.

The contributions of this work are as follows:
\begin{itemize}
\item A unified, modular mixed-integer linear programming (MILP) formulation that embeds three DC flexibility models (spatial, temporal, and joint spatio-temporal) within a single SCUC structure, enabling apples-to-apples comparison under identical generator, network, and \textit{N}-1 security constraints.
\item The marginal value of spatial and temporal flexibility is 
separately quantified within a security-constrained scheduling 
framework, showing that temporal flexibility dominates cost 
reduction while spatial flexibility better absorbs renewable 
curtailment, and that joint coordination captures synergistic 
benefits beyond either dimension alone.
\item A sensitivity analysis on the flexibility ratio $\beta$ identifies a diminishing-returns regime that yields a practical guideline: moderate flexibility ($\beta\!=\!0.20$ to $0.30$) captures most of the achievable grid relief.
\end{itemize}
\vspace{-5pt}
\section{System Model and Methodology}
\label{sec:methodology}

The proposed framework extends a standard SCUC formulation \cite{b12,b13} to incorporate flexible DC loads as controllable demand-side resources. The enhanced model co-optimizes generator commitment and dispatch, DC load allocation, and renewable generation utilization, subject to network flow, \textit{N}-1 contingency security, and DC energy conservation constraints. The objective function is linear in all decision variables and is given by:
\begin{align}
\min \; & \sum_{g,t} \Bigl(C^{\mathrm{E}}_g P_{g,t} + C^{\mathrm{NL}}_g u_{g,t} + C^{\mathrm{SU}}_g v_{g,t}\Bigr) \notag \\
       & + M\!\left( \sum_{l,t}\alpha_{l,t} + \sum_{l,c,t}\alpha^{c}_{l,t}
         + \sum_{n,t}\sigma_{n,t} + \sum_{n,c,t}\sigma^{c}_{n,t} \right)
\label{eq:obj}
\end{align}
where $C_g^{\rm E}$ is a constant linear energy cost coefficient. 
The first line minimizes total generation cost, while the second 
penalizes thermal violations and load shedding in both base-case 
and post-contingency states via a uniform large penalty $M$.

\textit{Generator dispatch and reserve} \cite{b12,b13}:
\begin{align}
P_g^{\min} u_{g,t} &\leq P_{g,t} \leq P_g^{\max} u_{g,t} - R_{g,t}^v, &\forall g,t \label{eq:gen_lb_ub}\\
R_{g,t}^v &\leq R_g^{\mathrm{SP}} u_{g,t}, &\forall g,t \label{eq:reserve_ub}\\
\sum_{g} R_{g,t}^v &\geq \max_{g \in \mathcal{G}}\{P_g^{\max}\}, &\forall t \label{eq:reserve_sys}
\end{align}
Constraint~\eqref{eq:gen_lb_ub} bounds dispatch within committed capacity net of spinning reserve. Constraints~\eqref{eq:reserve_ub}--\eqref{eq:reserve_sys} enforce individual and system-level reserve requirements, where $\max_{g \in \mathcal{G}}\{P_g^{\max}\}$ is a pre-computed constant equal to the rated capacity of the largest unit, ensuring reserve coverage under the loss of the single largest committed unit.

\textit{Start-up detection:}
\begin{align}
v_{g,t} \geq u_{g,t} - u_{g,t-1}, \quad \forall g,t \label{eq:startup}
\end{align}
where $u_{g,0} := u_{g,|\mathcal{T}|}$ enforces the cyclic horizon boundary.

\textit{Hourly ramp and minimum up/down time} (MILP form following \cite{b12,b13}):
\begin{align}
P_{g,t} - P_{g,t-1} &\leq R^{\mathrm{H}}_g u_{g,t-1} + R^{\mathrm{SR}}_g v_{g,t}, \;\;\forall g,t \label{eq:ramp_up}\\
P_{g,t-1} - P_{g,t} &\leq R^{\mathrm{H}}_g u_{g,t} + R^{\mathrm{SD}}_g (v_{g,t} \!-\! u_{g,t} \!+\! u_{g,t-1}), \;\;\forall g,t \label{eq:ramp_dn}\\
\sum_{\mathclap{s=t-T^{\mathrm{up}}_g+1}}^{t} v_{g,s} &\leq u_{g,t}, \quad \forall g,\; t \geq T^{\mathrm{up}}_g \label{eq:min_up}\\
\sum_{\mathclap{s=t+1}}^{t+T^{\mathrm{dn}}_g} v_{g,s} &\leq 1 - u_{g,t}, \quad \forall g,\; t \leq |\mathcal{T}|\!-\!T^{\mathrm{dn}}_g \label{eq:min_dn}
\end{align}
where $R^{\mathrm{SR}}_g$ and $R^{\mathrm{SD}}_g$ allow distinct ramping rate limits during start-up and shut-down, respectively.

\textit{Network flow and nodal balance:}
\begin{align}
P^{f}_{l,t} &= \frac{\theta_{f_l,t} - \theta_{t_l,t}}{x_l}, &\forall l,t \label{eq:dcpf}\\
-F^{A}_l - \alpha_{l,t} &\leq P^{f}_{l,t} \leq F^{A}_l + \alpha_{l,t}, \;\;\alpha_{l,t}\!\geq\!0, &\forall l,t \label{eq:thermal}
\end{align}
The reference-bus phase angle is fixed at zero. The thermal limit in~\eqref{eq:thermal} is enforced as a \emph{soft} constraint via a non-negative slack $\alpha_{l,t}$ priced at $M$ in~\eqref{eq:obj}. With $M$ set well above any per-MWh generation cost, the model recovers a hard limit whenever a feasible dispatch exists, while preserving feasibility under the inflexible-DC baseline and yielding an economic measure of network stress that is directly comparable across cases. The same rationale applies to $\sigma_{n,t}$ and to all post-contingency slacks.

The nodal power balance requires that generation, renewable injection, and net branch flows satisfy demand at each bus:
\begin{align}
&\sum_{g\in\mathcal{G}_n} P_{g,t} + R^{\mathrm{use}}_{n,t}
  - \sum_{l\in\delta^{+}(n)} P^{f}_{l,t}
  + \sum_{l\in\delta^{-}(n)} P^{f}_{l,t} \notag\\
&\quad = P^{d}_{n,t} + D_{n,t} - \sigma_{n,t}, \quad \forall n,t
\label{eq:pbal}
\end{align}
where $R^{\mathrm{use}}_{n,t}\!=\!0$ $\forall\, n\!\notin\!\mathcal{S}$ and $D_{n,t}\!=\!0$ $\forall\, n\!\notin\!\mathcal{D}$. Renewable utilization and load-shedding bounds are:
\begin{align}
R^{\mathrm{use}}_{n,t} + R^{\mathrm{curt}}_{n,t} &= R^{\mathrm{avail}}_{n,t}, &\forall n\in\mathcal{S},t \label{eq:pv_bal}\\
0 \leq \sigma_{n,t} &\leq P^{d}_{n,t} + D_{n,t}, &\forall n,t \label{eq:shed_ub}
\end{align}

\textit{\textit{N}-1 post-contingency security:}
For each contingency $c \in \mathcal{K}$ (single-branch outage), the following constraints are enforced. DC flexibility and renewable dispatch are optimized for the base case only; $R^{\mathrm{use}}_{n,t}$ and $D_{n,t}$ remain fixed across all post-contingency states. The outaged branch carries zero flow, while surviving branches follow the power flow equations with emergency thermal limits:
\begin{align}
P^{f,c}_{c,t} &= 0, &\forall c,t \label{eq:cont_pf_out}\\
P^{f,c}_{l,t} &= \frac{\theta^{c}_{f_l,t} - \theta^{c}_{t_l,t}}{x_l}, &\forall l \neq c,\; c,t \label{eq:cont_pf}\\
|P^{f,c}_{l,t}| &\leq F^{C}_l + \alpha^{c}_{l,t}, &\forall l \neq c,\; c,t \label{eq:cont_thermal}
\end{align}
Generators redispatch within their spinning reserve range while respecting output bounds:
\begin{align}
|P^{c}_{g,t} - P_{g,t}| &\leq R^{\mathrm{SP}}_g u_{g,t}, &\forall g,c,t \label{eq:cont_gen}\\
P^{\min}_g u_{g,t} &\leq P^{c}_{g,t} \leq P^{\max}_g u_{g,t}, &\forall g,c,t \label{eq:cont_gen_lim}
\end{align}
The post-contingency nodal balance mirrors~\eqref{eq:pbal} with re-dispatched outputs $P^{c}_{g,t}$ and possible load shedding $\sigma^{c}_{n,t}$:
\begin{align}
&\sum_{g\in\mathcal{G}_n} P^{c}_{g,t} + R^{\mathrm{use}}_{n,t}
  - \sum_{l\in\delta^{+}(n)} P^{f,c}_{l,t}
  + \sum_{l\in\delta^{-}(n)} P^{f,c}_{l,t} \notag\\
&\quad = P^{d}_{n,t} + D_{n,t} - \sigma^{c}_{n,t}, \quad \forall n,c,t
\label{eq:cont_pbal}
\end{align}
where post-contingency load shedding is bounded by:
\begin{align}
0 \leq \sigma^{c}_{n,t} &\leq P^{d}_{n,t} + D_{n,t}, \quad \forall n,c,t
\label{eq:shedc_ub}
\end{align}
\vspace{-25pt}
\subsection*{Data Center Load Decomposition and Flexibility Ratio $\beta$}
\label{sec:dc_flex}
\vspace{-3pt}
The total load of DC $n$ at time $t$, denoted $D_{n,t}$, is a decision variable co-optimized with unit commitment. It is decomposed as
\begin{equation}
D_{n,t} = D^{\mathrm{fix}}_{n,t} + D^{\mathrm{flex}}_{n,t}, \quad \forall n\in\mathcal{D},\; t\in\mathcal{T},
\label{eq:dc_decomp}
\end{equation}

where all DCs share a common normalized workload profile $D^{\mathrm{orig}}_t$. The inflexible component $D^{\mathrm{fix}}_{n,t}=(1-\beta)D^{\mathrm{orig}}_t$ represents non-shiftable services such as lighting and networking, and $\beta\in[0,1]$ is the flexibility ratio. The flexible energy budget declared by the DC operator is $\hat{D}^{\mathrm{flex}}_{n,t} = \beta\, D^{\mathrm{orig}}_t$.

Site capacity limits are enforced by:
\begin{align}
D^{\mathrm{fix}}_{n,t} \;\le\; D_{n,t} \;\le\; \overline{D}_n, \quad &\forall n\in\mathcal{D},\; t\in\mathcal{T},
\label{eq:dc_ub}\\
D^{\mathrm{flex}}_{n,t} \;\ge\; 0, \quad &\forall n\in\mathcal{D},\; t\in\mathcal{T}.
\label{eq:dc_nonneg}
\end{align}

Three flexibility models are formulated as variants of the same MILP, sharing a common objective~\eqref{eq:obj} and constraints~\eqref{eq:gen_lb_ub}--\eqref{eq:dc_nonneg}, and differing only in the energy-conservation constraint imposed on $D^{\mathrm{flex}}_{n,t}$, as summarized in Table~\ref{tab:models}: the Data Center Spatial model (DC-S), the Data Center Temporal model (DC-T), and the Data Center Spatio-Temporal model (DC-ST).
\vspace{-5pt}\vspace{-5pt}\vspace{-2pt}
\begin{table}[H]
\caption{Flexibility Model Formulations}
\label{tab:models}
\centering
\footnotesize
\setlength{\tabcolsep}{5pt}
\renewcommand{\arraystretch}{1.3}
\begin{tabular}{l c c c}
\toprule
\textbf{Model} & \textbf{Obj.} & \textbf{Shared Constr.} & \textbf{Unique Constr.} \\
\midrule
DC-S: Spatial         & \multirow{3}{*}{\eqref{eq:obj}} & \multirow{3}{*}{\eqref{eq:gen_lb_ub}--\eqref{eq:dc_nonneg}} & \eqref{eq:m1_spatial} \\
DC-T: Temporal        &  &  & \eqref{eq:m2_temporal} \\
DC-ST: Spatio-temporal &  &  & \eqref{eq:m3_spatiotemp} \\
\bottomrule
\end{tabular}
\end{table}
\vspace{-5pt}\vspace{-2pt}
\textit{Model-specific constraints:}
\begin{align}
  \sum_{n\in\mathcal{D}} D^{\mathrm{flex}}_{n,t}
    &= \sum_{n\in\mathcal{D}} \hat{D}^{\mathrm{flex}}_{n,t},
    &\forall\, t\in\mathcal{T}
    \label{eq:m1_spatial}\\
  \sum_{t\in\mathcal{T}} D^{\mathrm{flex}}_{n,t}
    &= \sum_{t\in\mathcal{T}} \hat{D}^{\mathrm{flex}}_{n,t},
    &\forall\, n\in\mathcal{D}
    \label{eq:m2_temporal}\\
  \sum_{n\in\mathcal{D}}\!\sum_{t\in\mathcal{T}} D^{\mathrm{flex}}_{n,t}
    &= \sum_{n\in\mathcal{D}}\!\sum_{t\in\mathcal{T}} \hat{D}^{\mathrm{flex}}_{n,t}
    \label{eq:m3_spatiotemp}
\end{align}

Constraint~\eqref{eq:m1_spatial} preserves system-wide instantaneous flexible demand at each time step, enabling cross-site workload migration without temporal shifting.
Constraint~\eqref{eq:m2_temporal} preserves each site's total flexible energy over the horizon, enabling intertemporal load shifting without cross-site transfer.
Constraint~\eqref{eq:m3_spatiotemp} enforces only the global energy budget, jointly permitting spatial and temporal reallocation. Since the feasible regions of DC-S and DC-T are both contained within that of DC-ST, the latter spans the largest feasible region and provides an upper bound on achievable system benefit.
\vspace{-3pt}
\section{Case Studies and Discussion}
\vspace{-5pt}
\subsection{Case Study Setup}
\vspace{-5pt}
The framework is evaluated on the IEEE 24-bus Reliability Test System (RTS-24): 24 buses, 38 branches, 33 units, and $|\mathcal{K}|=37$ \textit{N}-1 contingencies, with $\mathcal{D}=\mathcal{S}=\{9,18\}$. Two DC sites at buses 9 and 18 each follow a diurnal hyperscale profile (13,614~MWh/day; peak 720~MW, valley 443~MW) with site capacity $\overline{D}=850$~MW. A co-located 3,380~MW PV plant (30\%/70\% split between buses 9/18) is profiled from NREL Houston-region clear-sky data \cite{b15} and sized so that the Fixed-DC baseline exhibits binding network constraints and observable curtailment, enabling meaningful cross-case comparison. The flexibility ratio $\beta$ takes values in $\{0.10, 0.20, 0.30, 0.40\}$. Five cases are evaluated: Base (no DC), Fixed-DC, DC-S, DC-T, and DC-ST. All models are implemented in Pyomo and solved with Gurobi at a 1\% MIP gap.
\vspace{-6pt}
\subsection{System Cost and Penalty}
\vspace{-4pt}
Table~\ref{tab:summary} reports the total system objective and penalty for each case. The objective value includes both generation cost and the penalty for transmission violations and load shedding. The Base case incurs $\approx$\$516K with zero violations. Introducing concentrated, inflexible DC loads (Fixed-DC) drives the objective to \$7.31M, of which \$5.47M reflects transmission congestion and load-shedding penalties---a stark illustration of the network stress imposed by inflexible large-scale loads.
\vspace{-7pt}\vspace{-5pt}
\begin{table}[H]
\caption{System Objective and Penalty (\$, Millions)}
\label{tab:summary}
\centering
\vspace{-6pt}
\scriptsize
\renewcommand{\arraystretch}{0.95}
\setlength{\tabcolsep}{1.8pt}
\begin{tabular}{l | cc | cc | cc | cc}
\toprule
 & \multicolumn{2}{c|}{$\beta\!=\!0.10$}
 & \multicolumn{2}{c|}{$\beta\!=\!0.20$}
 & \multicolumn{2}{c|}{$\beta\!=\!0.30$}
 & \multicolumn{2}{c}{$\beta\!=\!0.40$} \\
\cmidrule(lr){2-3}\cmidrule(lr){4-5}\cmidrule(lr){6-7}\cmidrule(lr){8-9}
\textbf{Case} & Obj & Pen & Obj & Pen & Obj & Pen & Obj & Pen \\
\midrule
Base (no DC)           & 0.52 & 0.00 & 0.52 & 0.00 & 0.52 & 0.00 & 0.52 & 0.00 \\
Fixed-DC               & 7.31 & 5.47 & 7.31 & 5.47 & 7.31 & 5.47 & 7.31 & 5.47 \\
DC-S: Spatial          & 6.66 & 4.84 & 6.14 & 4.31 & 5.96 & 4.12 & 5.90 & 4.06 \\
DC-T: Temporal         & 3.92 & 2.02 & 3.31 & 1.39 & 3.12 & 1.17 & 3.04 & 1.09 \\
DC-ST: Spatio-temporal & 3.69 & 1.83 & 2.56 & 0.64 & 2.12 & 0.19 & 1.91 & 0.00 \\
\bottomrule
\end{tabular}
\vspace{2pt}
\begin{minipage}{\linewidth}
\footnotesize
\textit{Note:} $\beta$ is the fraction of $D_t^{\mathrm{orig}}$ available for deferral or spatial migration; $(1-\beta)D_t^{\mathrm{orig}}$ is served as scheduled. Obj: total system objective (generation cost plus penalty); Pen: penalty for thermal violations and load shedding (does \emph{not} include renewable curtailment, which is reported separately in Section~III-F).
\end{minipage}
\end{table}
\vspace{-10pt}
\vspace{-5pt}
All three models yield monotone improvement with $\beta$. The pronounced gap between DC-T and DC-S at every flexibility level indicates that temporal load mismatches---rather than spatial mismatches---constitute the dominant congestion driver in this network. DC-ST achieves zero penalty at $\beta\!=\!0.40$, confirming complete elimination of all base-case and post-contingency transmission violations through joint spatiotemporal coordination.
\vspace{-7pt}
\subsection{Data Center Load Profile Reshaping}
\begin{figure}[H]
\centerline{\includegraphics[width=\columnwidth]{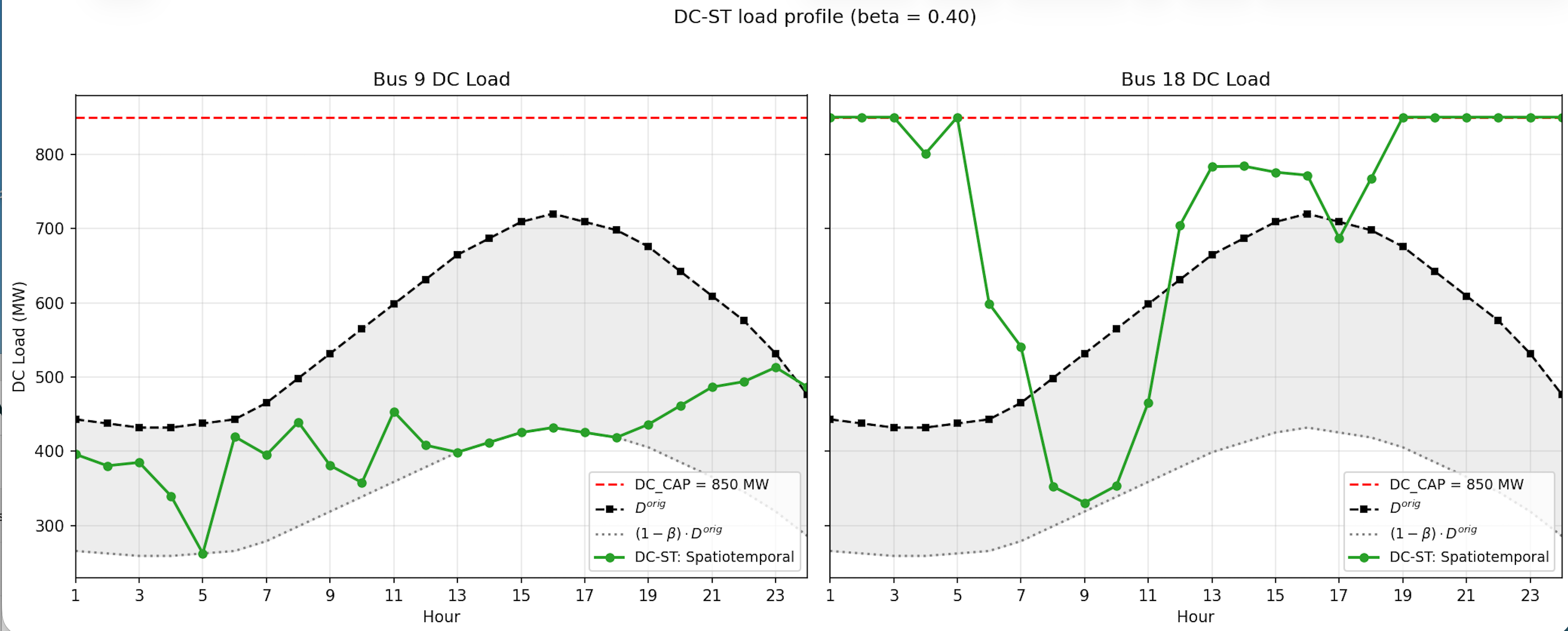}}
\vspace{-6pt}
\caption{DC-ST load profiles at Bus~9 (left) and Bus~18 (right) under $\beta=0.40$.
Dashed: original demand $D^{\mathrm{orig}}_t$; dotted: fixed portion $(1-\beta)D^{\mathrm{orig}}_t$;
solid green: optimized $D_{n,t}$ under spatio-temporal flexibility (DC-ST).
The red dashed line marks the 850~MW site capacity.}
\label{fig:dc_load}
\end{figure}

\vspace{-4pt}
Fig.~\ref{fig:dc_load} illustrates the 24-hour DC-ST load reshaping at the two co-located sites (Bus~9 and Bus~18) under $\beta\!=\!0.40$, revealing how joint spatio-temporal coordination redistributes workload both across sites and across time.
\vspace{-10pt}
\vspace{-10pt}
Under DC-ST, the two sites exhibit clearly complementary roles. Bus~9 operates as a net workload \emph{exporter}: its dispatched load $D_{n,t}$ stays well below $D^{\mathrm{orig}}_t$ throughout the day, dropping to the $(1-\beta)D^{\mathrm{orig}}_t$ floor near hour~5 and remaining near 400~MW thereafter, never approaching the 850~MW capacity ceiling. Bus~18, in contrast, acts as a net \emph{importer}, saturating at the 850~MW ceiling during the overnight valley (hours 1--5) and the evening peak (hours 19--24), while sharply offloading to the $(1-\beta)D^{\mathrm{orig}}_t$ floor during hours 7--10 to relieve morning ramp-induced congestion. The midday plateau at roughly 770--790~MW during hours 13--17 further coincides with the local PV peak, allowing concentrated absorption of renewable output. This dual reshaping---spatial migration from Bus~9 to Bus~18 combined with temporal redistribution at Bus~18 itself---is what enables DC-ST to simultaneously achieve the lowest operating cost and the largest curtailment reduction across all $\beta$ values, as quantified in the subsequent subsections.
\vspace{-8pt}
\subsection{Violation Reduction}
\begin{figure}[H]
\centering
\includegraphics[width=0.55\linewidth]{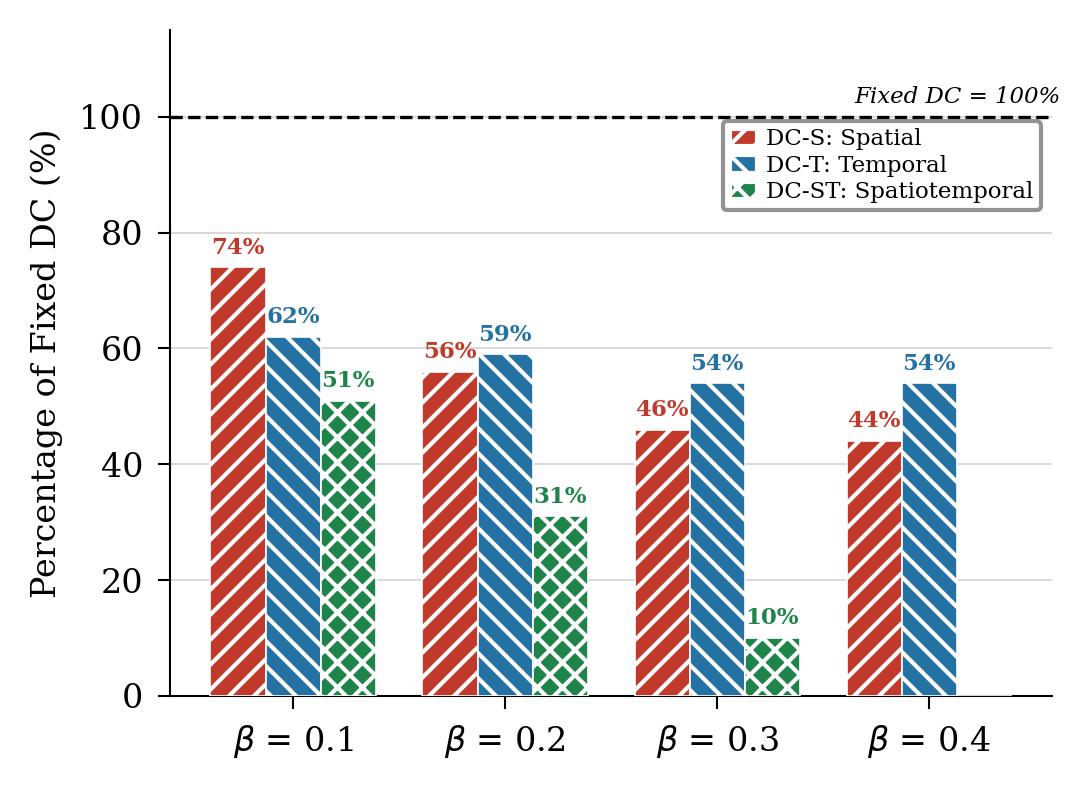}
\vspace{-5pt}\vspace{-5pt}\vspace{-2pt}
\caption{Number of lines with flow violations under contingency across flexibility models, normalized to Fixed DC ($= 100\%$).}
\label{fig:thermal}
\end{figure}

\vspace{-5pt}
Table~\ref{tab:violation} reports the percentage reduction in total constraint violations (thermal overloads plus load shedding, in both base case and post-contingency states) relative to Fixed DC; renewable curtailment is treated separately in Section~III-F. Fig.~\ref{fig:thermal} disaggregates the contingency thermal overload component, showing the number of lines with flow violations under contingency, normalized to the Fixed DC baseline.
\vspace{-7pt}\vspace{-5pt}
\begin{table}[H]
\caption{Constraint Violation Reduction vs.\ Fixed DC (\%)\\\footnotesize Thermal overloads + load shedding; distinct from renewable curtailment.}
\label{tab:violation}
\centering
\vspace{-4pt}\vspace{-2pt}\vspace{-2pt}
\footnotesize
\renewcommand{\arraystretch}{1.1}
\setlength{\tabcolsep}{3pt}
\begin{tabular}{l | c | c | c | c}
\toprule
\textbf{Case} & $\beta\!=\!0.10$ & $\beta\!=\!0.20$ & $\beta\!=\!0.30$ & $\beta\!=\!0.40$ \\
\midrule
DC-S: Spatial         & 11.6\% & 21.2\% & 24.8\% & 25.8\% \\
DC-T: Temporal        & 63.0\% & 74.6\% & 78.6\% & 80.0\% \\
DC-ST: Spatiotemporal & 66.5\% & 88.3\% & 96.5\% & 100.0\% \\
\bottomrule
\end{tabular}
\end{table}
\vspace{-5pt}
\vspace{-10pt}\vspace{-5pt}
\vspace{-2pt}
\vspace{-5pt}
\vspace{-5pt}
DC-S achieves only modest reduction across the tested range, reaching at most 25.8\% at $\beta\!=\!0.40$, confirming that purely spatial redistribution cannot resolve temporal imbalances: without intertemporal freedom, workload migration merely displaces rather than dissipates congestion. As shown in Fig.~\ref{fig:thermal}, DC-S still retains 44\% of the Fixed DC overload count at $\beta\!=\!0.40$, while DC-ST reduces it to 10\%. DC-ST eliminates 88.3\% of violations at $\beta\!=\!0.20$ and reaches full elimination at $\beta\!=\!0.40$.
\vspace{-9pt}
\subsection{Cost Savings and Sensitivity to $\beta$}
\begin{figure}[H]
\centering
\includegraphics[width=0.65\linewidth]{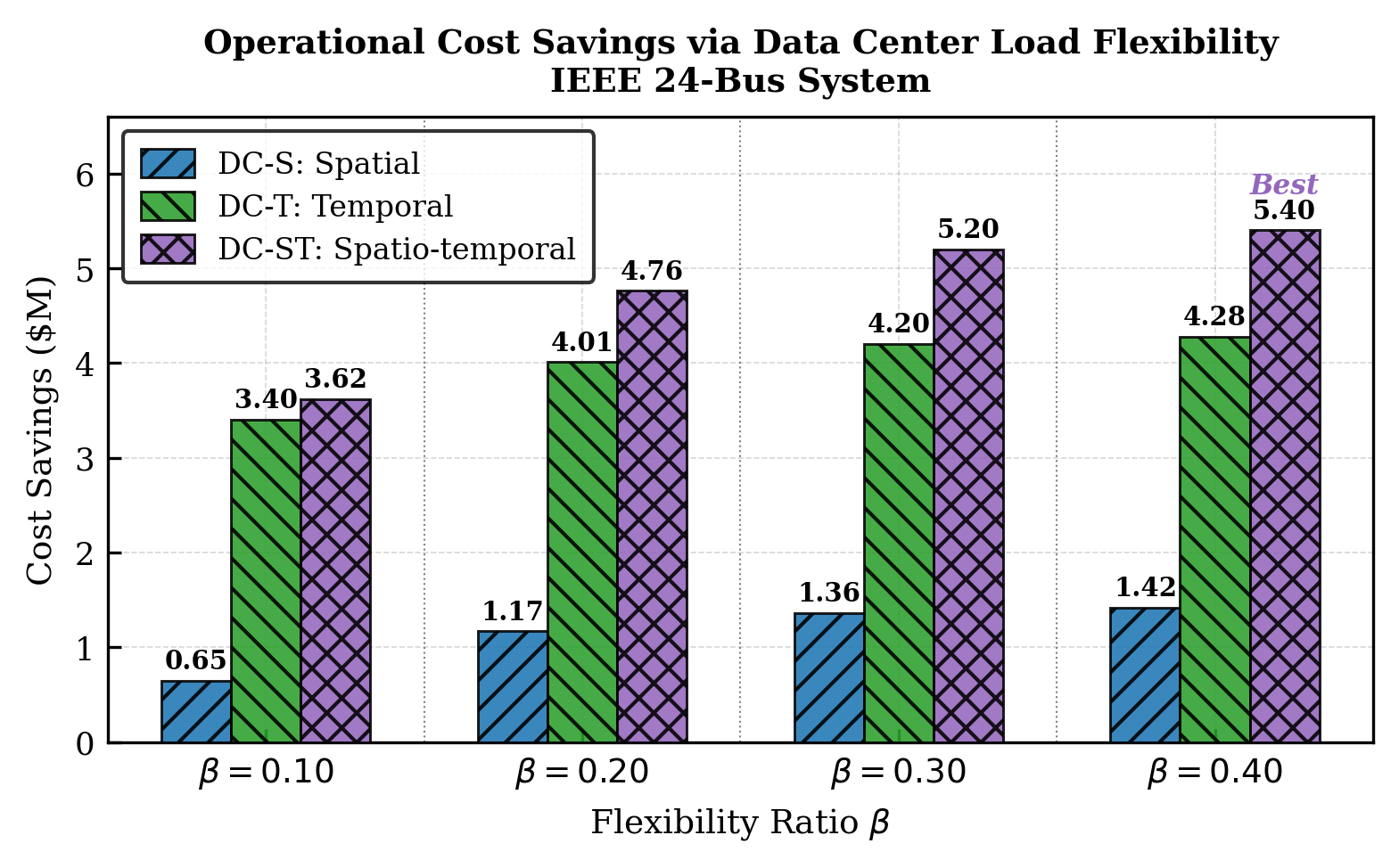}
\vspace{-5pt}\vspace{-5pt}
\caption{Cost savings vs.\ Fixed DC (\$M): DC-S (Spatial), DC-T (Temporal), DC-ST (Spatiotemporal) at $\beta \in \{0.10,\, 0.20,\, 0.30,\, 0.40\}$.}
\label{fig:1}
\end{figure}

\vspace{-5pt}
Fig.~\ref{fig:1} presents absolute cost savings relative to Fixed DC across $\beta$ values.
\vspace{-1pt}
DC-ST savings rise from \$3.62M at $\beta\!=\!0.10$ to \$5.40M at $\beta\!=\!0.40$, with the marginal gain per unit $\beta$ diminishing monotonically beyond $\beta\!=\!0.20$. The mechanism is intuitive: at small $\beta$, the limited flexible budget is preferentially deployed against the most binding constraints (the most overloaded lines and highest-priced congestion intervals), where each MWh of reallocated load relieves a large penalty; once these bottlenecks are eliminated, additional flexibility can only act on progressively less binding constraints, so each incremental $\Delta\beta$ buys less penalty reduction. This also accounts for the diminishing marginal cost savings observed across all three models as $\beta$ grows. Operationally, moderate contractual flexibility ($\beta\!=\!0.20$ to $0.30$) suffices to capture most of the achievable grid relief, substantially reducing the operational burden on DC operators.
\vspace{-5pt}\vspace{-5pt}
\vspace{-5pt}\vspace{-2pt}\vspace{-5pt}\vspace{-5pt}\vspace{-2pt}
\subsection{Renewable Curtailment Reduction}
\begin{figure}[H]
\centerline{\includegraphics[width=0.66\columnwidth]{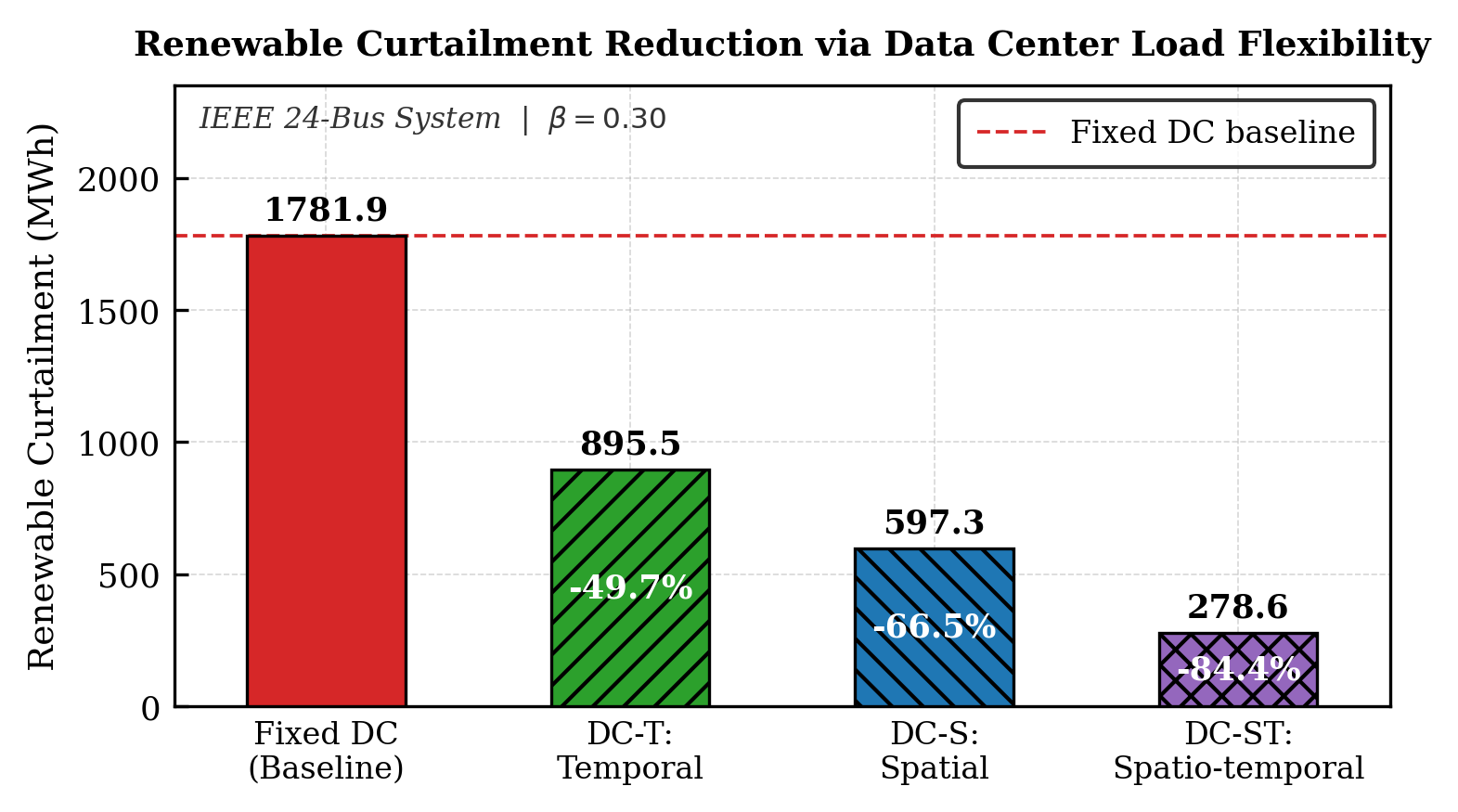}}
\vspace{-5pt}\vspace{-5pt}
\caption{Renewable curtailment at $\beta\!=\!0.30$: absolute (MWh) and reduction relative to Fixed DC (\%).}
\label{fig:curtailment}
\end{figure}

\vspace{-2pt}
Fig.~\ref{fig:curtailment} presents curtailment results at $\beta\!=\!0.30$. The Fixed DC baseline curtails 1,781.9~MWh over 24~hours. DC-T reduces curtailment to 895.5~MWh (49.7\%) by concentrating load during midday hours~9--16, when PV peaks and lines are otherwise overloaded. DC-S achieves 597.3~MWh (66.5\%) by directing workload toward the site with greater local renewable availability, cutting inter-regional power transfer. DC-ST attains 278.6~MWh, an 84.4\% reduction, by jointly exploiting both mechanisms. This curtailment metric is conceptually distinct from the constraint-violation reduction in Table~\ref{tab:violation}.
\vspace{-7pt}\vspace{-7pt}\vspace{-7pt}
\vspace{-7pt}\vspace{-5pt}
Notably, the ranking of DC-S and DC-T inverts between cost and curtailment metrics. DC-S is more effective at renewable absorption due to geographically heterogeneous renewable availability, while DC-T dominates congestion relief in the cost dimension. This metric-dependent inversion is topology- and resource-mix-specific, reinforcing the case for joint spatio-temporal coordination as a robustly superior strategy across varying grid configurations.
\vspace{-5pt}
\section{Conclusion}
\vspace{-5pt}
This paper presented a modular SCUC framework that integrates DC load flexibility as a system-level demand-side resource. Three MILP-encoded models (DC-S, DC-T, DC-ST) were evaluated on a modified IEEE 24-bus system with co-located DCs and PV. A clear hierarchy emerges: DC-ST delivers the highest benefit, eliminating all base-case and post-contingency violations at $\beta\!=\!0.40$ and cutting renewable curtailment by 84.4\% at $\beta\!=\!0.30$; DC-T captures most cost savings via temporal reshaping; DC-S attains at most 25.8\% violation reduction. The diminishing-returns behavior across all models indicates that moderate flexibility ($\beta\!=\!0.20$--$0.30$) suffices to capture most achievable grid relief. Future work will extend the framework to larger systems, incorporate renewable forecast uncertainty, and examine market mechanisms for DC participation.
\vspace{-5pt}\vspace{-5pt}
\section*{Acknowledgment}
\vspace{-5pt}
This research was supported in part by the University of Houston Energy Transition Institute Seed Grant Program.
\vspace{-5pt}
\bibliographystyle{IEEEtran}

\bibliography{references}
\vspace{-7pt}
\end{document}